\documentclass[final,5p,times,twocolumn]{elsarticle}
\usepackage{graphicx}
\usepackage{amssymb}
\usepackage{amsthm}
\usepackage{bm}
\usepackage{amsmath}
\newcommand{\beq}{\begin{equation}}
\newcommand{\eeq}{\end{equation}}
\newcommand{\beqa}{\begin{eqnarray}}
\newcommand{\eeqa}{\end{eqnarray}}

\journal{Solid State Communications}

\begin{document}

\begin{frontmatter}

\title{Vacancy induced zero energy modes in graphene stacks:
The case of ABC trilayer}
\author[label1]{Eduardo V. Castro}
\author[label2]{M. Pilar  L\'opez-Sancho}
\author[label2]{Mar\'{i}a A. H. Vozmediano}

\address[label1]{CFIF, Instituto Superior T\'{e}cnico, TU Lisbon, Av. Rovisco
Pais, 1049-001 Lisboa, Portugal }

\address[label2]{Instituto de Ciencia de Materiales de Madrid, Consejo Superior de
Investigaciones Cient{\'{\i}}ficas, Cantoblanco, 28049 Madrid, Spain }

\begin{abstract}

The zero energy modes induced by vacancies in ABC stacked trilayer
graphene are investigated. Depending on the position of the vacancy,
a new zero energy solution is realised, different from those obtained
in multilayer compounds with Bernal stacking. The electronic modification induced in the sample by 
the new vacancy states is characterised by computing the local density of states and 
their localisation properties are studied by the inverse 
participation ratio. We also analyse the situation  in the presence of a gap in
the spectrum due to a perpendicular electric field. 
\end{abstract}


\end{frontmatter}


\section{Introduction}
\label{intro}

Recent experimental advances aiming
to generate better samples for electronic devices have allowed
the obtention of high quality samples not only of monolayer 
graphene but also of bilayer graphene (BLG) and trilayer graphene (TLG)
\cite{Oetal08,CSetal09,KWH+12}.
One of the
major problems preventing applications of monolayer graphene is
the difficulty to open and control a gap in the samples.
To this
respect bilayer and multilayer samples are more promising
\cite{Oetal08}. 
The band structure of few layer graphenes  depends on the stacking order  \cite{APP10,BJetal11,LCetal11}, what offers the possibility of tuning
electronic properties. After the great excitement awaken by the bilayer compound due to the possibility of opening a tunable gap with an external gate \cite{MF06,CNM+06,Oetal08} the interest has moved recently to the trilayer materials due to their enigmatic properties. 
As in the Bernal AB stacked BLG an external electric field 
allows a tunable band gap in the ABC-stacked TLG \cite{KM09,ZSMMD10,Gap2}
while the ABA-stacked TLGs are semimetals with electric field tunable
band overlap \cite{CSetal09,JCetal11}.

Structural defects -- vacancies, ad-atoms, and other -- which may appear during the fabrication process are very important in the graphene materials. While in the first times after the synthesis  the concern was that they may deteriorate the performance of graphene-based devices, later tendencies point to their positive use in some applications, as they make it possible to tailor the local properties of graphene and to achieve new functionalities \cite{BKetal10}. Being one-atom thick the graphene materials are extremely sensitive to the presence of adsorbed atoms and molecules  and, more generally, to defects such as vacancies, holes and/or substitutional dopants. This property, apart from being directly usable in molecular sensor devices, can also be employed to tune graphene electronic properties. The possibility of a controlled manipulation of atoms and molecules on graphene has opened a new  area of research that allows to observe chemical interactions or structural modifications of low contrast molecules or nano-objects \cite{MGetal08}. Among the defect-induced properties in graphene, magnetism is one of the most appealing \cite{CGPNG09,Yaz10,LJV09,CSV11}. Recent experimental findings \cite{NSetal12} have ascertain the role of vacancies in the magnetic properties of the material.

Vacancies are recognised as important scattering centers in monolayer and BLG \cite{MOWFGB10}
and are known to induce zero-energy modes \cite{UBGG10} that  modify the low-energy
properties of the samples.
The existence and nature of localised states arising from
vacancies in BLG was analyzed in a recent paper
\cite{CLV10}. It was found that the two different types of
vacancies that can be present in the bilayer system 
give rise to two different
types of states: quasilocalised states, decaying as $1/r$ at
long distances $r$ from the vacancy, similar to these found in the monolayer case
\cite{PGS+06}, and truly delocalised states whose wave function
remains constant as $r\to\infty$. These later new midgap states 
were found to become localised inside the
gap  induced in the bilayer sample by the electric field
effect. The analysis of these vacancy-induced states was generalized
in \cite{CLV10} to multilayer graphene systems with ABAB$\cdots$ Bernal stacking.
The recent experimental and theoretical activity around the trilayer compounds 
\cite{ZZetal11,BJetal11,IRK11,GLM11,KEetal11}
has shown that the system in the rhombohedral
ABC stacking has very different
electronic properties than its Bernal partner. 

In this paper we will revise the situation of 
the midgap states induced by vacancies in 
few layers graphene, with especial attention to TLG.
We will see that the ABC-rhombohedral stacked  presents yet a new type
of zero-energy state with no analogue in the mono or bilayer compounds.
The characterisation of this new state exhausts the possibilities for the vacancy states
in multilayer graphene with the usual stacking.
The paper is organized as follows: in section \ref{sec:structure}
we give a summary of the situation encountered in  the bilayer system and
its extension to  multilayer graphene with Bernal (AB) stacking. 
Next we explain in section \ref{sec_ABC} the new
aspects encountered in the trilayer material with rhombohedral (ABC) stacking.
In section \ref{sec_end}
we summarise the analysis of the work and discuss possible
physical consequences.

\section{Vacancy states in Bernal stacked multilayer graphene }
\label{sec:structure}
The study and characterisation of vacancy states in monolayer graphene has been
an important subject that started prior to the synthesis of the material 
and continues  to our days \cite{CGPNG09,NSetal12} driven in part by the 
search for magnetism in $sp$ carbon compounds \cite{Yaz10}. In the monolayer case there is only one type
of vacancy whose presence induces two degenerate quasi-localised states at the Fermi energy. An analytic construction of the wave function of the vacancy was done in \cite{PGS+06} by matching surface state solutions at zigzag
edges with those localised at Klein edges for a suitable
boundary condition. In the continuum limit the wave function 
can be written as
\begin{equation}
\Psi(x,y)\approx\frac{e^{i\mathbf{K}.\mathbf{r}}}{x+iy}+
\frac{e^{i\mathbf{K}'.\mathbf{r}}}{x-iy},
\label{eq:PsiVac1L}
\end{equation}
where $\mathbf{K}$ and $\mathbf{K}'$ are the reciprocal space vectors
of the two inequivalent corners of the first Brillouin zone, and $(x,y)$
are distances in a reference frame centred at the vacancy position. The 
wave function is peaked at the position of the vacancy and
decays as $1/r$ away from it, a behaviour termed quasi-localised. 

This procedure to describe vacancy states was generalised to the AB stacked BLG
in \cite{CLV10}. Because of the similitudes with the trilayer case and to fix the notation we will present the main features of the bilayer in some detail in what follows.

\begin{figure}[t]
\begin{centering}
\includegraphics[width=0.99\columnwidth]{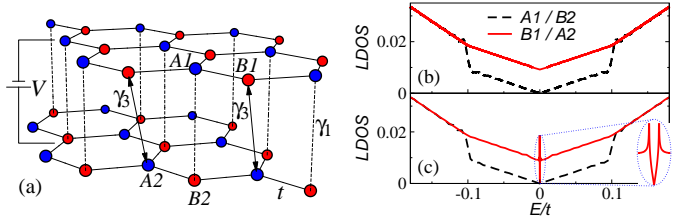}
\par\end{centering}
\caption{\label{fig:bilayer}(color online). (a)~Bilayer lattice structure
and main tight-binding parameters. (b)~The general shape of the density of states for the minimal model. The inset shows the changes in the DOS near the Fermi point when the various tight binding parameters are included.}
\end{figure}

The lattice structure of a BLG is shown in
Fig.~\ref{fig:bilayer}. In the 
AB-Bernal stacking  the top layer is shifted with respect to the bottom layer by one C--C distance. As a result only half of the atoms in any layer have a direct neighbour joined by $\gamma_1$ in the other layer. We use indices~1
and~2 to label the top and bottom layer, respectively. 
The main tight-binding hopping parameters are also shown in the figure as well as the different electronic structure near the Fermi level for different values of the tight binding parameters. In the minimal model adopted in most of the works only $\gamma_1$ is different from zero \cite{MF06}. The estimated values of the parameters are $t\sim 3eV$ and $\gamma_{1}/t\sim0.1$ \cite{CGPNG09}.

In the Bernal BLG there are two different types of vacancies giving
rise to unpaired atoms: vacancies
located at $A1/B2$ or $B1/A2$ {[}see Fig.~\ref{fig:bilayer}(a)].
The first type $A1/B2$ is produced by removing a site having a neighbour
in the adjacent layer and is usually named $\beta$ vacancy. The second type $B1/A2$ resulting
when the removed site is not connected to the other layer is called $\alpha$.

We obtained an  analytical solution for  the states induced by these 
two types of vacancies in \cite{CLV10} generalising the procedure done in 
the monolayer case \cite{PGS+06}. For vacancies at $A1/B2$ sites the wave function obtained is
the same given by Eq.~\eqref{eq:PsiVac1L} which
corresponds to  a quasilocalised, zero-energy
state decaying as $1/r$ around the vacancy in the same layer 
where the vacancy sits but in the opposite sublattice. This vacancy has then the same
properties as these found in the monolayer case. The new state found in \cite{CLV10}
corresponds to vacancies at $B1/A2$ lattice sites. The solution in this case has the form
\begin{equation}
\Upsilon(x,y)\sim
\left[\Psi(x,y),\,
\frac{\gamma_{1}}{t} \frac{x-iy}{x+iy} e^{i\mathbf{K}.\mathbf{r}}+
\frac{\gamma_{1}}{t} \frac{x+iy}{x-iy} e^{i\mathbf{K}'.\mathbf{r}}
\right],
\label{eq:Upsilon2L}
\end{equation}
 where $\Psi(x,y)$ is the quasi-localised state given in Eq.~(\ref{eq:PsiVac1L}),
and the two component wave function 
$\Upsilon \sim \left[\phi_1,\,\phi_2 \right]$ refers to the two layers; first
and second components for the first and second layers, respectively.
This is a delocalised state, with the peculiarity of being quasi-localised
in one layer (where the vacancy sits) and delocalised in the other
where it goes to a constant when $r\rightarrow\infty$.

The same analytical construction followed  for the minimal model 
in BLG can be directly applied to multilayer graphene in Bernal stacking
and for the particular case of the ABA TLG. The quasi-localised state~\eqref{eq:PsiVac1L}
is a solution in any multilayer with a $A1/B2$-vacancy. For a $B1/A2$-vacancy
the solution is a generalisation of state~\eqref{eq:Upsilon2L} with
a quasi-localised component in the layer where the vacancy resides
and delocalised components in the layers right on top and below this
one: $\Phi(x,y)\sim\left[\Psi(x,y),\,\Upsilon_2(x,y),\,\Upsilon_2(x,y)\right]$,
where $\Upsilon_2(x,y)$ refers to the second component in the right hand side
of Eq.~\eqref{eq:Upsilon2L}.

The behaviour of the vacancy states was ascertained in the bilayer system by numerical
calculations.  A summary of the findings of our previous work on Bernal stacked multilayer graphene is the following:
Associated to the presence of vacancies
and to the existence of a gap in the spectrum generated by an
electric field $E_z$ three different types of behaviour for vacancy-induced states are
found:
\begin{enumerate}
\item For $E_z=0$ a $\beta-$vacancy ($A1/B2$ site) induces  
quasi-localized state ($1/r$ behaviour).
\item For $E_z=0$ an $\alpha-$vacancy ($B1/A2$ site) induces a resonance
due to a delocalized state.
\item For  $E_z\neq 0$ a $\beta-$vacancy produces a resonance inside the
continuum near the band edge while an $\alpha-$vacancy gives rise to a
truly localized state inside the gap. This is the most interesting
state for the magnetic implications.
\end{enumerate}

\section{ABC trilayer}
\label{sec_ABC}

\subsection{Model}
\begin{figure}
  \centering
  \includegraphics[width=1.\columnwidth,clip]{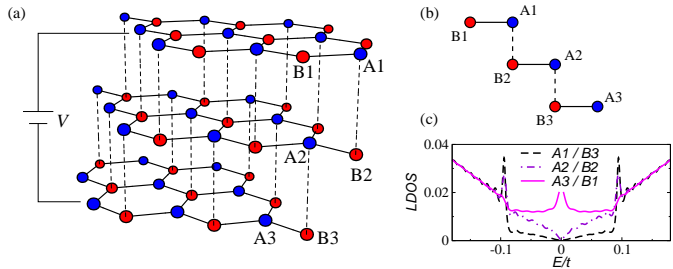} 
\caption{(color online) (a)~Trilayer lattice structure.
(b)~Scheme indicating in-plane (full line) and inter-layer (dashed line)
nearest neighbor hopping between different sublattices. (c)~LDOS at the 
three non-equivalent sites in perfect trilayer.}
\label{schemeLDOS}
\end{figure}
We follow the tight-binding approximation and consider
the minimal model where the in-plane hopping energy, $t$,
and the inter-layer hopping energy, $\gamma_{1}$, define the most
relevant energy scales. This is the minimal model for AB stacked
multilayer graphene described above. The simplest
tight-binding Hamiltonian describing non-interacting $\pi-$electrons
in TLG then reads:

\begin{equation}
H_{TB}=\sum_{i=1}^{3}H_{i}+\gamma_{1}\sum_{\mathbf{R},\sigma}
\big[a_{1,\sigma}^{\dagger}(\mathbf{R})b_{2,\sigma}(\mathbf{R})+
a_{2,\sigma}^{\dagger}(\mathbf{R})b_{3,\sigma}(\mathbf{R})+
\mbox{h.c.}\big],
\label{eq:Htrilayer}
\end{equation}
with $H_i$ being the SLG Hamiltonian
\begin{eqnarray}
H_{i}&=-t\sum_{\mathbf{R},\sigma}\big[a_{i,\sigma}^{\dagger}(\mathbf{R})
b_{i,\sigma}(\mathbf{R})+a_{i,\sigma}^{\dagger}(\mathbf{R})
b_{i,\sigma}(\mathbf{R}-\mathbf{a}_{1})+\\ \nonumber
& a_{i,\sigma}^{\dagger}(\mathbf{R})b_{i,\sigma}(\mathbf{R}-\mathbf{a}_{2})+
\mbox{h.c}\big],
\label{eq:Hslg}
\end{eqnarray}
where $a_{i,\sigma}(\mathbf{R})$ {[}$b_{i,\sigma}(\mathbf{R})$] is
the annihilation operator for electrons at position $\mathbf{R}$
in sublattice $Ai$ ($Bi$), $i=1,2,3$, and spin $\sigma$,
and $\mathbf{a}_{1}$ and $\mathbf{a}_{2}$ are the primitive vectors
of the underlaying Bravais lattice. If a perpendicular electric
field is applied \cite{Gap2}, the following potential energy should be added
to Eq.~\eqref{eq:Htrilayer},
\begin{equation}
H_{V}=\frac{V}{2}\sum_{\mathbf{R},\sigma}\big[n_{1,\sigma}(\mathbf{R}) - 
n_{3,\sigma}(\mathbf{R}\big],
\label{eq:V}
\end{equation}
with
$n_{i,\sigma}=a_{i,\sigma}^\dagger(\mathbf{R})a_{i,\sigma}(\mathbf{R})+
b_{i,\sigma}^{\dagger}(\mathbf{R})b_{i,\sigma}(\mathbf{R})$. 
We use the same values
 for in-plane and inter-layer hopping as for AB stacked multilayer graphene
 \cite{CGPNG09},
which, as we have seen above,
imply $\gamma_{1}/t\sim0.1\ll1$. Extra hopping
terms, as long as they preserve the bipartite nature of the
lattice, may introduce quantitative changes but not qualitative
\cite{PGS+06,CLV10,CLV09}. Hopping terms that break the bipartite
character of the lattice may be treated perturbatively afterwards
\cite{CSV11}.

The main feature of the ABC compound that makes it different from its Bernal ABA counterpart is the lack of mirror symmetry with respect to the middle layer. As discussed in \cite{GNP06,MGV07}  this type of  staking  allows the derivation of a low energy  effective hamiltonian  which  involves  only the unlinked  atoms $(A_1,B_3)$  given by
\beq
\mathcal{H}^{eff}=-\frac{v_F^3}{\gamma_1^{2}}\left(
\begin{array}{cc}
0 & k^{*3}\\
k^{3} & 0
\end{array}
\right),
\label{eq_Hcontinuum}
\eeq
where $k=k_x+ik_y$ and $v_F = 3ta/2$, with $a$ the C--C distance.
It also guarantees the  topological stability of the low-energy chiral effective Hamiltonian and makes its behavior similar to the AB bilayer discussed previously \cite{MGV07}. As it happens in the bilayer case, a gap can be induced in the ABC system by an external gate. 

Vacancies are modelled as missing sites in Eq.~\eqref{eq:Htrilayer}.
There are three non-equivalent vacancy sites. As can be inferred from
the lattice sketch shown in Fig.~\ref{schemeLDOS}(b), these sites 
correspond to a vacancy occurring in sublattice 
$A1/B3$, $A2/B2$, and $A3/B1$. The LDOS at these three lattice sites
is shown in Fig.~\ref{schemeLDOS}(c) for the perfect lattice
(no vacancy present). Despite the oscillatory behaviour due to finite
size effects, distinct thermodynamic limit features can be seen \cite{GNP06}.
The low energy physics is determined by sublattices $A3/B1$, and at
$E = 0$ a Van-Hove singularity develops due to the cubic spectrum $E\sim k^3$.
Sublattices $A1/B3$ and $A2/B2$ have a vanishing contribution to the
density of states at low energy. Due to interlayer hybridization 
their contribution is appreciable only for $|E| > \gamma_1$.

\subsection{New midgap state: Analytical considerations}
\label{sec_anal}
From what is known for the single layer and for bernal stacked multilayer
graphene we may immediately infer the following.
\begin{enumerate}
\item For a vacancy at any of the two possible places in the middle layer
of the ABC TLG the zero energy mode is of the new bilayer type:
quasi-localised in the middle layer, and delocalised in the adjacent
layer not connected to the vacancy.
\item For a vacancy at any of the outer layers, and residing in the sublattice
which is connected to the middle layer, then we simply have the single
layer solution: a state quasi-localised around the vacancy.
\item For a vacancy at any of the outer layers, but residing in the sublattice
which is not connected to the middle layer we have a new solution
with amplitude over the three layers simultaneously.
\end{enumerate}
By extension of the bilayer case we expect the new solution to have
a quasi-localised component in the layer with the vacancy and a delocalised
one in the middle layer. This is nothing but the new bilayer solution
for these two components. The correct behaviour of the third component
may be determined by the matching procedure used previously \cite{PGS+06,CLV10}. We do
not follow such prescription here. Instead, we perform a numerical
analysis detailed in the next section. Nevertheless, we may expect
this new component to be delocalised as well just based on the continuum,
low energy model,  Eq.~\eqref{eq_Hcontinuum} that describes ABC TLG. Far from the vacancy
position such component must satisfy $\partial_{z}^{3}\psi(z,\bar{z})=0$.
As a natural extension of the bilayer case we have the solution $\psi(z,\bar{z})=z^{2}f(\bar{z})$,
with $f(\bar{z})$ meromorphic. Assuming, as in the single layer and
bilayer cases, that the meromorphic  function is the usual $f(\bar{z})=1/\bar{z}$,
the new component is is manifestly delocalised.

\subsection{Numerical analysis}
\label{sec_numerical}
\begin{figure}
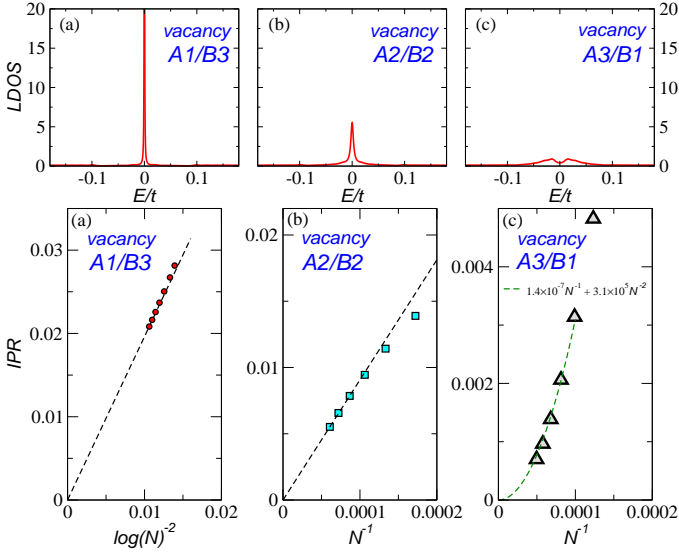

  \centering
  \includegraphics[width=1.\columnwidth,clip]{ldos_1Vac.V-0.eps} 
  \includegraphics[width=1.\columnwidth,clip]{ipr.V-0.eps} 
\caption{(color online) Upper: LDOS for a vacancy at sublattice (a) $A1/B3$, (b) $A2/B2$,
and (c) $A3/B1$. The LDOS is computed at a lattice site closest to the
vacancy. Down: IPR for the zero-energy mode induced by the previous vacancies.  In panels~(a) and~(b) linear guide lines are shown to  illustrate the scaling. In panel~(c) the line is a fit with linear and  quadratic terms.
}
\label{ldos1Vac}
\end{figure}
%

As mentioned before, vacancies are introduced in the TLG lattice by elimination of atom sites.
In our approach we use a simple tight binding Hamiltonian and do not include any reconstruction in the remaining structure.
We have analysed the changes induced
in the LDOS for sites around each  vacancy for the three different types described.
The LDOS is computed by 
the recursive Green's function method in clusters with $N=3\times1200^{2}$,
from which the thermodynamic limit can be inferred.

The localisation character of
vacancy-induced modes is studied through finite-size-scaling of the
inverse participation ratio (IPR). The later is defined as 
\beq
\mathcal{P}_{\nu}=\sum_{i}^{N}|\varphi_{\nu}(i)|^{4}
\eeq
for the eigenstate $\nu$, where $\varphi_{\nu}(i)$ is its amplitude
at site $i$. We perform exact diagonalization on small clusters with
$N$ up to $3\times82^{2}$ sites. The IPR for \emph{extended}, \emph{quasi-localized},
and truly \emph{localized} states scales distinctively with $N$ \citep{PdSN07}.
While for extended states we have $\mathcal{P}_{\nu}\sim N^{-1}$,
for quasi-localized states the $1/r$ decay implies $\mathcal{P}_{\nu}\sim\log(N)^{-2}$
(consequence of the definition of the IPR in terms of normalized eigenstates).
For localized wavefunctions the significant contribution to $\mathcal{P}_{\nu}$
comes from the sites in which they lie, and a size independent $\mathcal{P}_{\nu}$
shows up.

\subsubsection{Gapless case}

The LDOS and corresponding IPR are shown in Fig.~\ref{ldos1Vac} for the three types
of vacancy states discussed in this work. The LDOS shown in the upper part 
is computed at a lattice site closest to the vacancy. We can see a sharp resonant peak
for the first type of  vacancy in panel \ref{ldos1Vac}(a) that corresponds to the  state common to the monolayer.
The wave function has amplitude only in the layer of the vacancy 
and is quasi--localised   in the  opposite sublattice.
The $1/r$ decay of the wave function is apparent from the IPR shown in the down part \ref{ldos1Vac}(a). 

The middle panel \ref{ldos1Vac}(b) depicts the results obtained for the vacancy A2/B2. This type of vacancy,
also existing in the bilayer AB, is quasi-localised
in one layer (where the vacancy sits) and delocalised in the adjacent layer
where it remains constant when $r\rightarrow\infty$.
This behaviour can be seen as a broader feature in the LDOS (upper part \ref{ldos1Vac}(b)) that can be 
attributed to the quasi-localized component in the layer where the
vacancy sits (and thus the feature). This interpretation is fully
corroborated by the IPR scaling analysis shown in the down part \ref{ldos1Vac}(b).

Panel \ref{ldos1Vac}(c) shows the behaviour of the new vacancy state found in this work. The absence of a distinct resonance around zero energy is a clear
indication that the wave function of this vacancy has amplitude in all three layers. The feature associated with the quasi--localised component in the layer where the vacancy resides is thus weaker than in the other cases, and is burried in the continuum background of band states.
The results for the IPR 
(lower panel) \ref{ldos1Vac}(a) confirm the behaviour predicted
by the general arguments done in Sec.~\ref{sec_anal}.
The IPR scales as $N^{-1}$ as corresponds to an extended state,
even though finite size effects are quite strong in this case
(higher powers of $N^{-1}$ are necessary to fit the data).

\subsubsection{Gapped case}
\begin{figure}
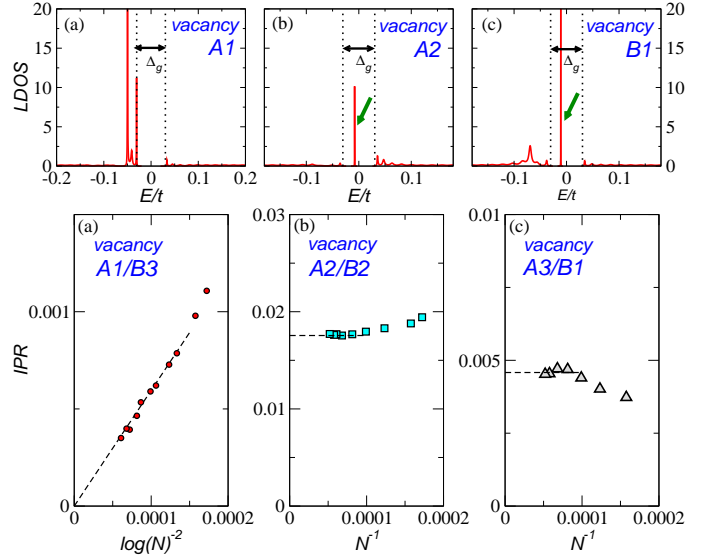

\label{gapldos1Vac}
  \centering
  \includegraphics[width=1.\columnwidth,clip]{ldos_1Vac.V-0.1.eps} 
  \includegraphics[width=1.\columnwidth,clip]{ipr.V-0.1.eps} 
\caption{(color online) Upper: LDOS for a vacancy at sublattice (a) $A1/B3$, (b) $A2/B2$,
and (c) $A3/B1$ for a finite gap, $V=0.1t$. The LDOS is computed at a 
lattice site closest to the vacancy. Localized modes inside the gap
are signaled by the arrow. Down: IPR for a vacancy at sublattice (a) $A1/B3$, (b) $A2/B2$,
and (c) $A3/B1$ for a finite gap, $V=0.1t$. As usually done in regions of continuum DOS, the IPR in~(a) is an average over states in an energy bin $\Delta E = 0.3t$ around the gap-edge resonance shown in 
the upper part~(a). In~(b) and~(c) the IPR is for the in-gap
mode shown in the upper part. Lines are guides to the eyes.}
\end{figure}
%

The ABC TLG presents, as the AB BLG, the possibility of opening and controlling a gap
in the spectrum by applying an external electric field $E_{z}$. We have studied the
behaviour of the midgap states when a gap opens. 
We consider a gap induced through a perpendicular electric
field $E_{z}=V/(ed)$, where $d\approx0.34\,\mbox{nm}$ is the interlayer
distance. Its presence is modelled by adding an on-site energy term: $-V/2$ at
layer~1 and $V/2$ at layer~3, as given by Eq.~\eqref{eq:V}.

The results for the gapped case are shown in Fig.~\ref{gapldos1Vac}. 
We plot the LDOS (upper panel) and IPR (lower panel)
for the three different types of vacancies. 

The first 
type of vacancy state \ref{gapldos1Vac}(a),
corresponds to the monolayer type having amplitude only
in the layer where the vacancy sits. From the LDOS it can be seen that the quasi-localised
state shown in the upper panel of Fig.~\ref{ldos1Vac}(a),  becomes a resonance
around $\pm V/2$ in the gaped case. The 
"monolayer" vacancy \ref{ldos1Vac}(a) becomes delocalised at the gap edge. The
IPR for this case is an average over the gap-edge resonance shown in the LDOS.

The LDOS for the vacancies of the types \ref{ldos1Vac}(b) and \ref{ldos1Vac}(c) (upper panel)
show that they live inside the gap. We have ascertained this result  performing
calculations for different gap sizes. Their asymmetric weight over the
two layers explains why they appear off zero-energy.
The IPR scaling to a constant
(lower panel) shows that these vacancies
give rise to truly localised states inside the gap. 
This is the same behaviour
encountered for the bilayer case in \cite{CLV10}.

\section{Conclusions and discussion}
\label{sec_end}
We have analysed the various midgap states that arise from the presence of vacancies in multilayer graphene with special emphasis on their degree of localisation.  Previous works based on a generalisation of the construction of vacancy states in bilayer AB to  multilayer compounds with Bernal (AB) staking found very different localisation properties for the two different types of vacancies that are present in these compounds. We have completed these works by analysing the rombohedral ABC trilayer graphene and found yet another new type of vacancy states.
Based in tight binding arguments ane can show that the new vacancy state survives beyond three layers. In the rhombohedral multilayer the amplitude spreads over all layers, though decaying exponentially  as $(\gamma_1/t)^n$ away from the layer where the vacancy sits ($n=0$).
We can also apply continuum arguments.
It was shown in \cite{MGV07} that the general rombohedral staking including the links
  $(A_1-B_2, A_2-B_3,\ldots, A_{N-1}-B_N)$  admits a low energy  effective hamiltonian
  which  involves  only the unlinked  atoms $(B_1,A_N)$,  given by
 \beq\label{eq_Hmul}
\mathcal{H}^{eff}  \sim -\frac{(t/a)^N}{ \gamma_1^{N-1}}\left(
\begin{array}{cc}
0 & k^{*N}\\
k^{N} & 0
 \end{array}
 \right)
\eeq
and hence the new  vacancy state found in the ABC trilayer for the $B_1$--$A_3$ is directly generalizable to 
$B_1$--$A_N$ in the multilayer rombohedral compounds.
 This finding exhausts the possibilities for the types of vacancy states in multilayer graphenes in the two more common stacking. 

A very interesting open issue is  the fate of these states for the twisted multilayers \cite{LPC07,BHetal12} although it is probable that the described features will persist for the vacancies having the same neighbours in the twisted superlattice as the ones described here.

The magnetic behaviour associated to the two types of vacancies of the bilayer system
was analysed in \cite{CLV09}. The fully localised midgap states arising from the new 
vacancy states in the presence of a gap will give rise to fully localised magnetic moments 
that will play a predominant role in the magnetic behaviour of the samples. 

\section{Acknowledgments}
Support from the Spanish Ministry of
Science and Innovation (MICINN) through  grants
PIB2010BZ-00512 and FIS2011-23713  is acknowledged.

\bibliographystyle{model1a-num-names}
\bibliography{Tril}

\end{document}